\documentclass{article}
\usepackage[xdvi]{epsfig}
\begin{document}

\title{The properties of dark matter}

\author{Ernst Fischer\\e.fischer.stolberg@t-online.de}
\maketitle


\begin{abstract}
Observations of density profiles of galaxies and clusters
constrain the properties of dark matter. Formation of stable halos
by collisional fluids with very low mass particles appears as the
most probable interpretation, while halos formed by high mass
particles, left over from a hot big bang, can scarcely explain the
observed density distributions. Detection methods of dark matter
are discussed.
\end{abstract}

\section {Introduction}
The fact that gravitational dynamics of the universe is dominated
by some matter component, which cannot be detected by any
electromagnetic signals, is one of the most discussed problems in
cosmology. There are two very pronounced effects, which are
ascribed to this dark matter. One is the observed high velocity of
galaxies in clusters, which exceeds by far the escape velocity of
the cluster, if only the masses of stars and gas are taken into
account. The other is the constancy of the rotation velocity of
spiral galaxies in the outer regions, where observed stars and gas
contribute only little to the matter budget.

Various ideas have been discussed to explain, what this dark matter may be,
beginning with weakly interacting particles, left over from a hot beginning
of the universe (WIMPs), new kinds of neutrinos or other very light
particles, proposed by some theories of grand unification (GUT) or
supersymmetric partners to our known particles, proposed by string theory.
Even the influence of particles living in extra dimensions has been invoked.
Rather comprehensive overviews on these ideas have been given by Ostriker and
Steinhardt \cite{ostriker} and more recently by Feng \cite{feng}.

Most of the papers in the field are inspired by concepts, which
relate the existence of dark matter to effects, which took place
in a very hot beginning of the universe and thus to the physical
processes, which may have occurred during the cooling process from
a state of extreme energy density. Depending on the physical model
of the 'big bang' process and on the underlaying models of high
energy particle physics, various particle candidates are discussed
and investigated with respect to their ability to act as dark
matter.

But all these theoretical models are more or less speculative and we have no
definite proofs of their correctness. Thus in this paper we will not follow
the usual way, but focus first on the question, what we can learn from
present observations on the properties of the dark matter particles and only
after that look for possible candidates, to meet these requirements. From the
fact that any good theory of particle physics must supply us with the
corresponding particles we can finally decide on the best choice or even
require a revision of all existing models.

\section{Properties of dark matter}
From observations of galaxy clusters only rather vague conclusions can be
drawn on the properties of dark matter particles. The fact that apart from
their influence on the gravitational balance we see no measurable effects is
only suitable to exclude possible properties. The most obvious statement,
which one can make, is that dark matter does non interact
electromagnetically. It has no electrical charge and cannot decay into
charged particle pairs or photons. Such events would be detectible as some
characteristic form of radiation.

The same argument holds for interactions with ordinary matter.
Collisions with neutral atoms would lead to excitation or
ionization, if the collisional energy is high enough. The absence
of such processes excludes dark matter particles with high
collisional energy, be it nonrelativistic particles with very high
mass or relativistic particles with low mass. These arguments have
led many authors to the conclusion that dark matter is
'collisionless'. But this can only refer to inelastic or reactive
collisions.

Elastic interactions with energy transfer far below the excitation
levels of atoms or molecules are possible, of course, as well as
interactions between dark matter particles. These interactions may
be direct collisions or momentum exchange in the gravitational
field of neighboring particles. Without an exchange of energy and
momentum dark matter particles could never be captured by existing
matter concentrations. They would be gravitationally accelerated
towards the center and after the passage would fly out again
without leaving any trace.

Thus observation of the rotation curves of spiral galaxies adds an
additional constraint to the properties of dark matter particles.
The constancy of the rotation velocity in the outer regions, where
dark matter dominates the mass distribution, requires a matter
density profile decreasing with distance from the center as
$1/r^2$. The radial dependence of the dark matter density can only
be understood, if these particles transfer energy and momentum
between one another, so that the energy which they take up from
the gravitational field can be transferred from radial into
thermal motion.

These constraints show that dark matter behaves just like an ideal
gas under the influence of gravitation, but the kinetic energy of
the individual dark matter particles must be less than the
excitation energy of ordinary atoms or molecules. Such an ideal
gas of low mass particles and with low energy density would
scarcely be able to condense into closed structures like galaxies
or galaxy clusters, as the energy gained from the gravitational
field would build up a pressure, which would counteract further
concentration. Instead fluctuations would develop into a web-like
structure as we observe it in the large scale structure of the
universe and which can be well reproduced by numerical
simulations. But in contrast, ordinary matter which can lose
energy by electromagnetic radiation, may form very concentrated
structures by gravitational instability, which then can attract
dark matter from an ubiquitous homogeneous distribution. By a
self-enhancing process this dark matter concentrates into a state,
where it dominates over ordinary matter.

\section{Modelling dark matter distributions}
The fact that dark matter halos are found as well around galaxy clusters as
around individual galaxies, be it small or large, be it old or young, must be
regarded as a hint that these halos are stable equilibrium configurations, at
least on a scale of galaxy life times, and not relics of an early epoch of
the universe. Thus as a first approximation we can regard them as assemblies
of some kind of ideal gas, which is in virial equilibrium. At least in the
outer regions the influence of ordinary matter to the equilibrium conditions
can be neglected, though condensations of ordinary matter should be regarded
as the primary source of structure formation.

Apparently dark matter particles feel no other force than
gravitation and can exchange kinetic energy or momentum only by
elastic gravitational interaction or by direct collisions. But as
will be discussed later, the contribution of distant interactions
appears negligible, as the mean free path is too large to
establish equilibrium in galaxy sized objects. The energy which
the particles take up in a gravitational field is transferred to
thermal motion preferably by collisions. Whether this thermal
energy remains in the region, where it is produced, or if it is
transported by conduction, depends on the mean free path. In the
limit of small mean free path the local energy density equals the
potential energy taken up during infall into the halo, the normal
virial condition
\begin{equation}
2E+U=2E_0+U_0,
\end{equation}
where $E$ and $U$ are the density of kinetic and potential energy and $E_0$
and $U_0$ the respective values outside the halo. Equilibrium is obtained
when the local energy is redistributed homogeneously into all degrees of
freedom yielding an homogeneous pressure $p=-U/3$. The dynamic equilibrium
condition in a radially symmetric halo can by described by
\begin{equation}
\label{pressure}
\frac{dp}{dr}=-\varrho\frac{d\Phi}{dr},
\end{equation}
\begin{figure}[hbt]\epsfig{file=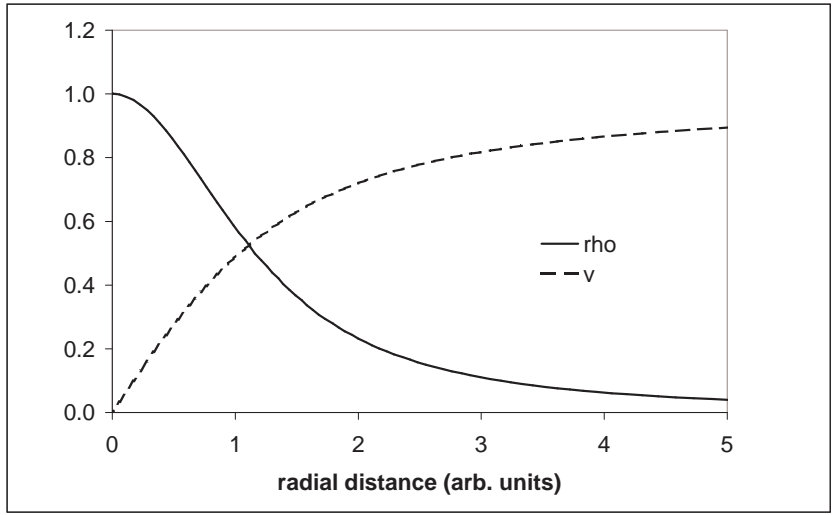,angle=-90,width=11cm}
\caption{Solid line: relative radial density distribution,
dashed line: circular equilibrium velocity}
\end{figure}
where $\Phi$ is the gravitational potential, defined in such a way that it
vanishes outside the halo. Under the assumption that ordinary matter with
mass $M_0$ is concentrated inside some radius $r_0$, outside this radius the
potential is given by
\begin{equation}
\label{mass}
\Phi-\Phi_0=\frac{G}{r}\left(M_0+4\pi \int_{r_0}^r \varrho\; r'^2dr'\right),
\end{equation}
where $\Phi_0$ has to be chosen so that $\Phi=0$ at infinity, assuming that
the dark matter has zero energy outside the halo. Thus with the ideal gas
condition $p=2/3 E$ and $U=-\varrho\Phi$ eq.(\ref{pressure}) reads
\begin{equation}
\frac{dp}{dr}=-\frac{1}{3}\frac{d(\varrho\,\Phi)}{dr}=-\varrho\frac{d\Phi}{dr}.
\end{equation}
with the general solution $\varrho= C\Phi^2 $. Together with the
relation between the potential and the matter density
eq.(\ref{mass}) this leads to the result that the potential
approaches zero as $1/r$ and the matter density decreases as
$1/r^2$. In the regions, where the density of ordinary matter is
negligible, the normalized radial profile takes the simple
analytical form
\begin{equation}
\varrho =(\tanh(r)/r)^2,
\end{equation}
which is valid up to the region, where the influence of other
structures becomes important or where the time of formation is too
long to reach equilibrium. Fig.1 shows the general form of the
density profile together with the expected profile of the
equilibrium circular velocity $v$, defined by $\varrho
v^2/r=\varrho \frac{d\Phi}{dr}$. The absolute value of the dark
matter density remains undefined from this model. It is determined
by the total amount available in the corresponding region of
space, which will be distributed among the ordinary matter
concentrations like galaxies during their period of formation.

It should be stressed here that the density profile of dark matter is quite
different from the frequently used NFW profile (Navarro, Frenk and White
\cite{NFW}), but agrees much better with observations, as well with respect
to the asymptotic power law $1/r^2$ as to the flat core distribution.

By now we have demonstrated that the most probable candidate of
dark matter is an ideal gas, which permeates all space and
condenses onto structures, which are formed by ordinary matter.
The dark matter particles do not exhibit any electromagnetic
interaction, but exchange energy and momentum only by direct
collisions and by gravity.

We can obtain further information on the nature of these particles
from the fact that, while collisions between them lead to
thermalization on the scale of galaxies, particles like protons or
nuclei, which we observe as cosmic rays, transverse large parts or
even the entire galaxy without considerable loss of energy. That
means that the number of collisions must be high, but the energy
transfer to protons or other nuclei is negligible. The mass of the
dark matter particles must be very small compared to the proton
mass.

It has been argued that dark matter should be collisionless
(Markevitch et al.\cite{mark}), because observations of two
merging galaxy clusters, the so called bullet cluster, show that
in determinations of the total matter distribution the dark matter
appears to follow the collisionless motion of galaxies, but the
hot x-ray emitting cluster gas is displaced due to ram pressure.
But this does only show that the dark matter is coupled to the
individual galaxies like a fixed atmosphere, unlike the
intracluster plasma, which appears continuously distributed all
over the cluster volume. Every galaxy is dressed by its own dark
matter halo. Thus the different behavior of the dark matter and
the hot plasma can be regarded as an additional hint that dark
matter is highly collisional and thermalized, so that it follows
the motions of concentrations of ordinary matter in galaxies.

We can try to get an idea on the mass of dark matter particles
from the fact that the halos are thermalized on the scale of
galaxies, that means that the mean free path of the particles is
much less than the characteristic length of the matter
distribution.

That distant gravitational interactions are negligible, can easily
be estimated. We consider the motion of a particle with initial
velocity $v$ in a medium of density $\varrho$ consisting of
particles of mass $m$, which is deflected by the gravitational
attraction of other particles. We put the question, how far it
must travel until a series of such small deflections adds up to a
change of the direction of motion, so that the perpendicular
velocity equals the the initial velocity $v_\bot=v$. The number
density of particles is $n=\varrho/m$, their mean distance
$d=1/n^{1/3}$. The impact parameter $b$ determines the deflection
of the particle trajectory. The force perpendicular to $v$ at
distance $r=\sqrt{x^2+b^2}$ is
\begin{equation}
F=\frac{Gm^2b}{r^3}
\end{equation}
(G is the gravity constant). Integrating from infinity to the
point of closest approach leads to a transverse velocity $
v_\perp=F/2m$ (both particles are deflected in opposite
directions)
\begin{equation}
v_\perp=\int_{-\infty}^b\frac{Gmb}{2vr^2\sqrt{r^2-b^2}}dr=\frac{Gm}{2bv}
\end{equation}
Averaging over all possible impact parameters from 0 to $d/2$
yields
\begin{equation}
v_\perp=\frac{4}{\pi d^2}\int_{0}^{d/2}\frac{Gm}{2bv}2\pi b\,
db=\frac{2Gm}{dv}
\end{equation}
The average perpendicular velocity after N deflections  of
statistical direction is $\frac{2Gm}{dv}\sqrt{N}$. The mean free
path $\lambda$ is reached when $v_\perp=v$.
\begin{equation}
\lambda=\frac{Nd}{2}=\frac{1}{2}\sqrt{\frac{d^3v^2}{2Gm}}=\sqrt{\frac{v^2}{2Gmn}}
=\sqrt{\frac{v^2}{2G\varrho}}
\end{equation}
independent of the size of the individual particles.

In a galaxy, where the density profile can be approximated by
$\varrho=A/r^2$ and the equilibrium condition $v^2/r=GM/r^2$
holds, this leads to the invariant condition
\begin{equation}
\lambda=\sqrt{\frac{\pi}{2}}r
\end{equation}
That means that the mean free path is in the order of the
characteristic length of the matter distribution. This condition
is based on the same physics as underlying the Jeans criterion of
gravitational instability. This gravitational interaction may be
responsible for the formation of large scale structure in the
universe, but to establish thermal equilibrium in the interior of
galaxy halos, much shorter mean free path is required.

There must be additional interactions between dark matter
particles. Interaction is not restricted to the mutual
gravitational attraction, but there must be also direct
collisions. As we take for sure that dark matter particles obey
some form of quantum physics, they must be regarded as fields with
a spatial extension of their wave function in the order of the
Compton wavelength. Consequently we can expect that there exist
interactions, when the distance between particles is less than
this length scale. The collision cross section thus is of the
order
\begin{equation}
\sigma =\left(\frac{hc}{\varepsilon}\right)^2
\end{equation}
where $\varepsilon$ is the rest energy of the particles. The number density
of particles in a region of total energy density $E$ is $n=E/\varepsilon$, so
that we obtain the mean free path
\begin{equation}
\label{lambda}
\lambda =\frac{1}{\sigma n}=\frac{\varepsilon ^3}{(hc)^2E}.
\end{equation}
In a galaxy this length must be small compared to that of the
gravitational interactions mentioned before. Let us take the
conditions in the solar system as a typical example. The rest
energy density can be calculated from the rotation velocity of the
galaxy in our neighborhood $v=220\; \rm{km/s}$. Assuming a
spherical matter distribution and a total density profile with
slope $1/r^2$, at the distance $r_0=6.7\;\rm{kpc}$ from the center
we find an energy density
\begin{equation}
E=\frac{v^2c^2}{4\pi Gr_0^2}=0.76\; \rm{GeV}.
\end{equation}
More detailed density profiles, taking additionally into account
stellar matter and gas distribution, give similar results, all in
the order of 1 GeV. The upper limit of the mean free path, to
achieve local equilibrium must be small compared to the local
characteristic length of the density gradient, that means, at most
of the order $\lambda=1\;\rm{kpc}$. Thus from eq.(\ref{lambda}) we
find $\varepsilon<36\;\rm{MeV}$. This rules out as constituents of
dark matter any of the GeV or TeV particles, proposed from
supersymmetry or similar elementary particle models.

Unfortunately we cannot derive any lower bound from such
estimations. The only thing we can learn from eq.(\ref{lambda}) is
that the mean free path changes with the third power of the
particle rest energy. Energies in the order of a few eV, as they
are discussed as rest energies of neutrinos, would result in a
mean free path in the order of cm. This would also rule out
neutrinos as possible candidates, as we observe neutrinos
travelling deep into the earth without considerable attenuation.
But we must keep in mind that our estimations are valid only for
thermal particles. Relativistic particles may exhibit interaction
cross sections many orders of magnitude lower.

There may exist other particle species, which are unknown to us by
now, though they exist everywhere in the solar system and even on
earth. But due to their extremely weak interaction with ordinary
matter they have escaped detection by now. If these particles
interact only gravitationally, they do not even 'know' what
electromagnetism is. Thus interactions with photons are excluded,
and even annihilation processes with the corresponding
antiparticles would be impossible, as there is no resulting photon
pair to carry the energy away. Such particles may exist as a
stable mixture of particles and antiparticles (or they are their
own antiparticles). The only way to change their number would be
annihilation by three body interactions and pair creation by
collisions in the extreme tail of their kinetic energy
distribution.

\section{Detection of dark matter particles}
From the properties of dark matter particles, which we have
discussed in the last section, it appears difficult to detect
individual particles directly. We expect that due to their low
mass and low kinetic energy reactive collisions with ordinary
matter are extremely improbable. All the experiments focussed to
the detection of GeV or TeV particles will scarcely obtain
positive results. They may find some high energy events. But these
should not be attributed to the dark matter, which forms the halos
around galaxies. Instead they must be caused by processes in
active galaxy cores or similar high energy environments.

As we assume that the kinetic energy of dark matter particles in
our environment is too low to react with ordinary matter, the only
way to obtain measurable interactions is, to supply the collision
energy by highly accelerated ordinary matter. There are sufficient
numbers of dark particles everywhere around us. From the rest
energy density of about $1 \,\rm GeV/cm^3$ in the solar
neighborhood we expect $10^3$ particles per $\rm cm^3$ with 1 MeV
rest energy or $10^9$ particles of 1eV.

Thus the best chance to observe reactions of these particles is in
accelerators like Tevatron or LHC, and it may well be that we
already have many such observations without considering them as
influenced by dark matter particles. Collisions in these
accelerators take place not in absolute vacuum, but in a medium
filled with the ubiquitous dark matter, which permeates the vacuum
tubes unimpeded. Thus collisions between the accelerated ordinary
matter particles may by chance take place in the presence of a
dark particle, which can change the balance of quantum numbers.
Observed violations of CP invariance in weak interactions or
flavor non-conservation may well be the result of interference
with the dark matter particles.

It has to be proved, if a systematic investigation of these
processes, which violate established quantum rules, can be
explained by interactions with dark matter. There is some hope
that it might be possible to extract additional information from
archived data, when they are examined in the light of this new
interpretation.

Of course, these ideas are somewhat speculative, but we should
keep them as a possible alternative, at least as we have no
observations conflicting with this concept. By now we have no
positive identification of dark matter particles, but from
observations we can rule out with high probability all the models,
which are based on some high mass particles, left over from a hot
big bang.

\end{document}